\documentclass[twocolumn,epsfig,APS,floats,showpacs]{revtex4}
\usepackage{graphicx}
\usepackage{color}

\usepackage{fancyhdr}
\fancypagestyle{titlepage}{
\fancyhead[R]{Journal of  Physics: Condensed Matter \textbf{27}, 415501 (2015)}}
\begin{document}

\title{The origin of anisotropy and high density of states in the electronic structure of
Cr$_2$GeC by means of polarized soft X-ray spectroscopy and \textit{ab initio} calculations}

\author{Martin Magnuson$^{1}$, Maurizio Mattesini$^{2,3}$, Matthieu Bugnet$^{4,5}$ and Per Eklund$^{1,4}$}

\affiliation{$^1$Department of Physics, Chemistry and Biology, IFM, Thin Film Physics Division, Link\"{o}ping University, SE-58183 Link\"{o}ping, Sweden}

\affiliation{$^2$Departamento de F\'isica de la Tierra, Astronom\'ia y Astrof\'isica I, Universidad Complutense de Madrid, Madrid, E-28040, Spain}

\affiliation{$^3$Instituto de Geociencias (CSIC-UCM), Facultad de CC. F\'isicas, E-28040 Madrid, Spain}

\affiliation{$^4$D\'epartement de Physique et Mecanique des Mat\'eriaux, Institut Pprime, UPR 3346 CNRS - Universit\'e de Poitiers - ENSMA, SP2MI, T\'el\'eport2, 86962 Futuroscope, France}

\affiliation{$^5$Department of Materials Science and Engineering, McMaster University, Hamilton, Ontario L8S 4L7, Canada.}

\date{\today}

\begin{abstract}

The anisotropy in the electronic structure of the inherently nanolaminated ternary phase Cr$_{2}$GeC is investigated by bulk-sensitive and element selective soft x-ray absorption/emission spectroscopy. The angle-resolved absorption/emission measurements reveal differences between the in-plane and out-of-plane bonding at the (0001) interfaces of Cr$_{2}$GeC. The Cr $L_{2,3}$, C $K$, and Ge $M_{1}$, $M_{2,3}$ emission spectra are interpreted with first-principles density-functional theory (DFT) including core-to-valence dipole transition matrix elements. For the Ge $4s$ states, the x-ray emission measurements reveal two orders of magnitude higher intensity at the Fermi level than DFT within the General Gradient Approximation (GGA) predicts. We provide direct evidence of anisotropy in the electronic structure and the orbital occupation that should affect the thermal expansion coefficient and transport properties. As shown in this work, hybridization and redistribution of intensity from the shallow $3d$ core levels to the $4s$ valence band explain the large Ge density of states at the Fermi level. 

\end{abstract}

\maketitle

\section{Introduction}
The group of ternary nanolaminated carbides and nitrides known as $M_{n+1}AX_{n}$-phases is the subject of intense research \cite{Wang,Eklund1}. Three related crystal structures are classified by stoichiometry as 211 (n=1), 312 (n=2) and 413 (n=3) phases, where the letter $M$ is an early transition metal, $A$ is an element in the groups III-V and $X$ is either carbon or nitrogen. The $M_{n+1}AX_{n}$-phases have a technologically important combination of properties that are metallic and ceramic \cite{Barsoum1}. These properties are related to the crystal structure, the choice of the three constituent elements, as well as the strength of the chemical bonding between the nanolaminated layers.

For the 211 type of crystal structure, there are more than 50 different ternary carbides and nitrides, where Cr$_{2}$GeC was recognized by Jeitschko \textit{et al.} already in the 1960s \cite{Jeitschko}. This multifunctional metallic and ceramic compound is particularly interesting as it exhibits a number of peculiar properties and is relatively little studied. First, Cr$_{2}$GeC has the highest thermal expansion coefficient $\alpha$, among all the presently known $M_{n+1}AX_{n}$-phases \cite{Bouhemadou,Scabarozi,Cabioch}, which is largest in the M-A bonds and smallest in the M-X bonds as indicated by thermal motions of the atoms in neutron diffraction experiments \cite{Nina}. 
There is also a discrepancy between the theoretical and experimental $\alpha$-values of 30\% \cite{Zhou}. 
This could be related to a change in the orbital occupation as the temperature is raised that allows the system to show large volume variation with temperature. 

Second, the magnetic ordering and electron correlations affect other properties such as thermal expansion and bulk modulus of Cr-based MAX-phases, even though they are only weakly magnetic \cite{Ramzan1,Ramzan2,Maurizio}. Although pure Cr$_{2}$GeC is non-magnetic macroscopically at room temperature, the individual Cr atoms in Cr$_{2}$GeC have a magnetic order that could depend on the temperature \cite{Dahlqvist,Jaouen}. However, due to Cr $3d$ electron correlation effects, the electronic origin of the unique properties is not well understood or conclusive. Transport properties cannot be interpreted in terms of ground state electronic structure calculations of Cr-based M$_{n+1}$AX$_{n}$-phase materials since the electron-phonon coupling is expected to be strong and most likely anisotropic.

Third, the calculated density of states (DOS) at $E_F$ (7.7 states/eV/unit cell for U=0, and 7.9 states/eV/cell with U$_{eff}$=2.09 eV ($t_{2g}$) and 2.33 eV ($e_{g}$) for a chosen GGA$^{WC}_{xc}$ functional) \cite{Maurizio}, dominated by the Cr $3d$ states, is much lower than the experimentally reported value (21 states/eV/cell \cite{Barsoum2} and 22 states/eV/cell \cite{Drulis}), and there is a large electron-phonon coupling \cite{Nina}. For Cr$_{2}$GeC, the reported value of the DOS at $E_F$ is by far the highest measured among the $M_{n+1}AX_{n}$-phases (Cr$_{2}$AlC has the second highest) \cite{Barsoum2}. 

\begin{figure}
\includegraphics[width=87mm]{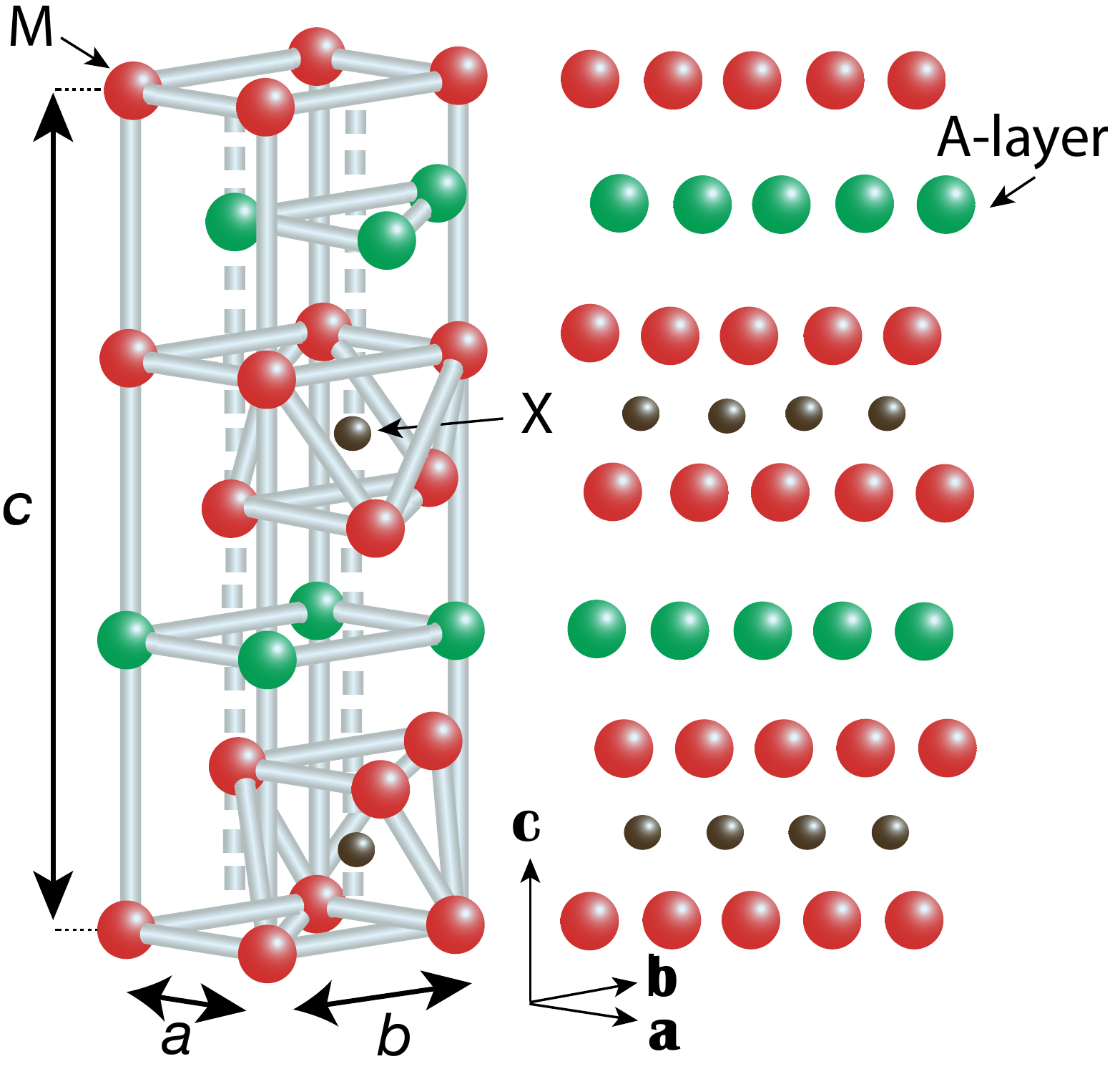} 
\vspace{0.2cm} 
\caption[] {(Color online) Illustration of the hexagonal crystal structure of Cr$_2$GeC. M is an early transition metal, A is an element in the groups III-V and X is either C or N. Every fourth layer of the CrC slabs are interleaved by pure layers of Ge. Irradiation by x-rays of specific energy and incidence angle from a synchrotron yields information about the occupied and unoccupied electronic orbitals of the chemical bonds $across$ ($d_{3z^2-r^2}$, $p_{z}$) and $in$ ($d_{xy}$, $d_{x^2-y^2}$, $p_{xy}$) the laminate plane. }
\label{structure}
\end{figure}

In this paper, we investigate the anisotropy and the orbital occupation in the electronic structure of single-crystal Cr$_{2}$GeC (0001) thin films. By applying bulk-sensitive and element-specific soft x-ray absorption (XAS), x-ray emission spectroscopy (XES) and resonant inelastic x-ray scattering (RIXS), we characterize the unoccupied and occupied bands of the containing elements, respectively \cite{Magnuson4,Magnuson5,Magnuson6}. This enables exploring the orbital occupation of the electrons buried several hundred nanometers below the surface. The energy of the x-rays is tuned to the specific core levels of Cr, Ge, and C and the incidence angle of the x-rays is changed from grazing to near normal relative to the laminate plane. This allows probing the occupation of the occupied electronic orbitals \textit{across} ($d_{3z^2-r^2}$, $p_z$) and \textit{in} ($d_{x^2-y^2}$, $d_{xy}$, $p_{xy}$) the laminate plane as illustrated in Figure 1. We provide direct evidence of anisotropy in the electronic structure supported by \textit{ab initio} band structure calculations. We further observe that the intensity ratio between the Ge $4s$ and $3p$ states in the valence band obtained from DFT-GGA is highly underestimated in comparison to the one measured via XES spectra.

\section{Experimental and computational details}

\subsection{Cr$_{2}$GeC (0001) thin film synthesis}
Cr$_{2}$GeC (0001) thin films were deposited by dc magnetron sputtering (base pressure $\sim$ 5x10\textsuperscript{-10} Torr) from elemental targets of Cr, C, and Ge in an argon discharge pressure of 4 mTorr on MgO (111) substrates. Details on the synthesis process are given in Ref.  \cite{Eklund2}. 

\subsection{X-ray emission and absorption measurements}
The XAS and XES measurements were performed at room temperature (300 K) at the undulator beamline 
I511-3 at MAX II (MAX-IV Laboratory, Lund, Sweden), comprising a 49-pole undulator, 
and a modified SX-700 plane grating monochromator \cite{MAX-lab,Nanocomp}. The measurements were made 
at a base pressure lower than 6.7*10\textsuperscript{-7} Pa. The XAS measurements were performed in total fluorescence yield (TFY) mode with 0.14 eV and 0.05 eV energy resolutions at the Cr $2p$ and C $1s$ absorption edges, respectively. 
The XAS-TFY spectra were measured at 15\textsuperscript{o} (along the $c$-axis, near perpendicular to the basal \textit{ab} plane) and 90\textsuperscript{o} (normal, parallel to the basal \textit{ab}-plane) incidence angles. 
The XAS spectra were normalized to the step before, and after the absorption edges and the self-absorption effects were estimated
with the program XANDA \cite{Klementev}.

The XES spectra were measured at 15\textsuperscript{o} (in the basal \textit{ab} plane) and 75\textsuperscript{o} (perpendicular to the basal \textit{ab}-plane) incidence angles. For comparison of the spectral profiles, the measured XES data were normalized to unity and plotted on a common photon energy scale (top) and relative to the $E_{F}$ (bottom). The XES spectra were recorded with 0.7, 0.2, 0.2 and 0.3 eV beamline monochromator energy resolutions at the Cr $L$, C $K$, Ge $M_{1}$, and Ge $M_{2,3}$ edges and 0.68, 0.19, 0.19 and 0.30 eV spectrometer resolutions, respectively. Self-absorption is known to affect the shape of the XES spectra on the high-energy flank of the main peak at the overlap with the XAS spectra. The observed XES intensity $I$, can be written as \cite{Cu}; 

\begin{equation}
I=I_o  \times \frac{1}{\left[ 1+\frac{\mu_{out}}{\mu_{in}} \times \frac{\sin\alpha}{\cos\beta  }  \right]  }   
\end{equation}

where I$_o$ is the unperturbed intensity, $\mu_{in}$ and $\mu_{out}$ are the absorption coefficients for the incident and outgoing radiation, $\alpha$ and $\beta$ are the incident and exit angles of the photon beam relative to the sample surface and surface normal, respectively. The estimated self-absorption effect is largest (5-10\% at E$_F$ depending on the excitation energy) when XAS is measured at grazing incidence (unoccupied out-of-plane states) and for XES at near normal incidence (occupied out-of-plane states). On the contrary, the effect is almost negligible (0-3\% at E$_F$ depending on the excitation energy) for XAS measured at normal incidence (unoccupied in-plane states) and XES measured at grazing incidence (occupied in-plane states).

\begin{figure}
\includegraphics[width=87mm]{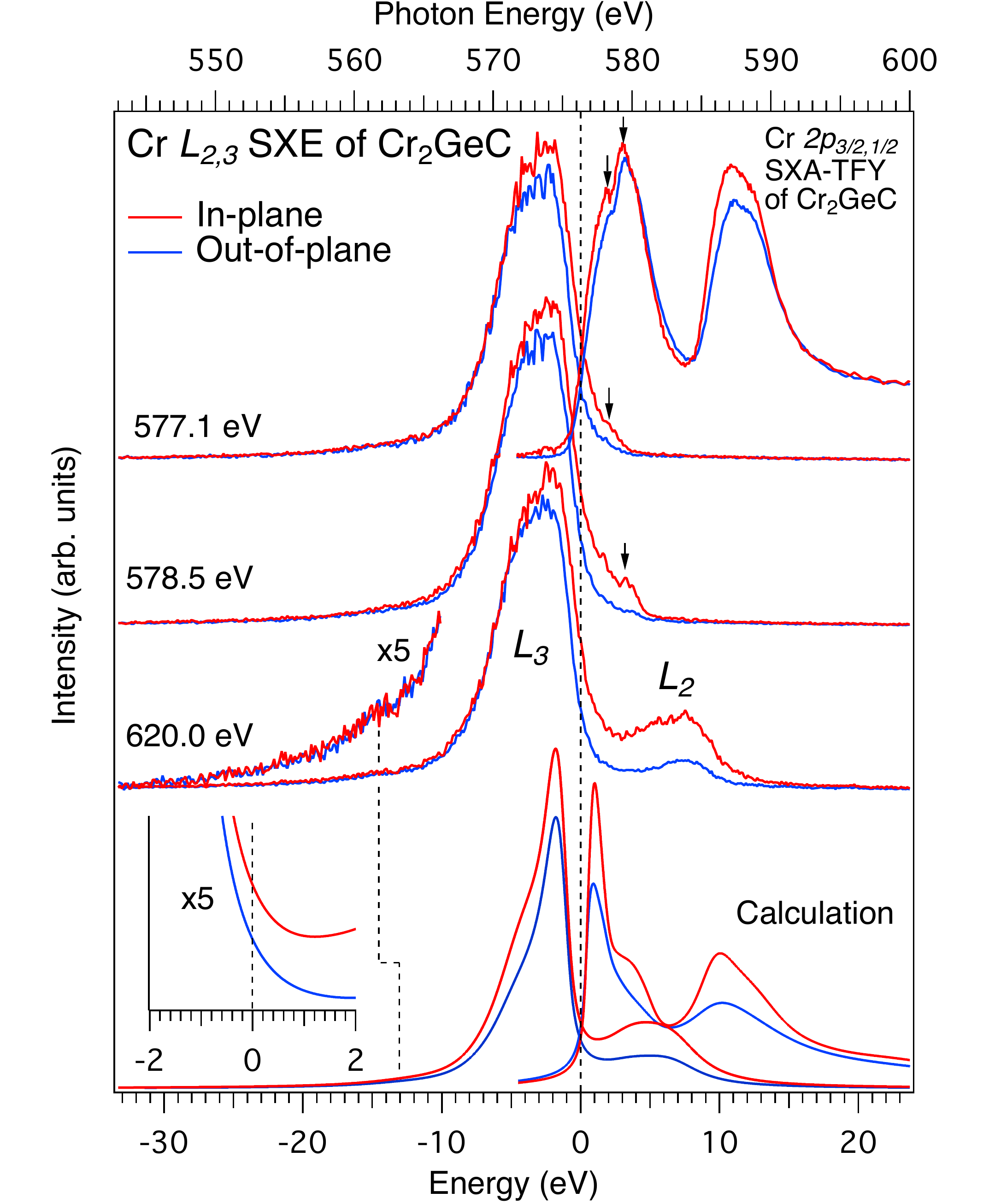} 
\vspace{0.2cm} 
\caption[] {(Color online) Top-right: experimental Cr $2p$ XAS-TFY spectra of Cr$_{2}$GeC following the $2p_{3/2,1/2} \rightarrow 3d$ dipole transitions measured at 15$^{o}$ 
(near perpendicular to the laminate plane) and 90$^{o}$ at normal incidence (parallel to the laminate plane). Center top-to-bottom: resonant Cr $L_{2,3}$ XES spectra with excitation energies 577.1, 578.5 eV, (indicated by the arrows in the XAS spectra), and nonresonant spectra at 620.0 eV. All spectra are plotted on a photon energy scale (top) and a relative energy scale (bottom) with respect to the top of the valence band. Bottom, calculated spectra using the experimental spin-orbit peak-splitting and $L_3$/$L_2$ branching ratio. }
\label{Cr 2p}
\end{figure}

\subsection{First-principles calculations}
All the \textit{ab initio} calculations performed in this work are based on the well-known Density Functional Theory (DFT) \cite{Kohn}. For the computations of the XES spectra, we used the so-called final-state rule \cite{vonBarth}, where no core-hole was created at the photo-excited atom. Particularly, emission spectra were obtained by means of the single-particle transition model using the Full Potential Linearized Augmented Plane Wave (FPLAPW) method \cite{Blaha}. Exchange and correlation effects were taken into account through the Generalized Gradient Approximation (GGA) as parameterized by Perdew, Burke and Ernzerhof (PBE) \cite{Perdew} using a matrix-size of R$_{MT}$*K$_{max}$=10, where K$_{max}$ is the plane wave cut-off and R$_{MT}$ the smallest of all atomic sphere radii (1.55 for carbon atoms). The charge density and potentials were expanded up to \textit{l}=12 inside the atomic spheres, and the total energy was converged with respect to the Brillouin zone (BZ) integration. The electric-dipole approximation was here used to calculate XES spectra, implying that only the electronic transitions arising from core levels with orbital angular momentum \textit{l} to the \textit{l}$\pm$1 components of the valence bands were studied. For core-hole lifetime and broadening parameters we used the available experimental numbers. In the fitting procedure to the nonresonant Cr $L_{2,3}$ XES spectra, we used the experimental values for the $L_3$/$L_2$ branching ratio of 10:1 in the basal plane and 15:1 along the \textit{c}-axis. For the XAS spectra, we used the experimental values for  $2p_{3/2}$/$2p_{1/2}$ branching ratio (1.1:1) and for the $L_{2,3}$ peak splitting (8.0 eV), which is somewhat larger than our calculated \textit{ab initio} spin-orbit splitting of 7.6 eV. Absorption spectra were calculated within the same theoretical scheme used for XES, but with the addition of core-hole effects \cite{LaCoO3}. Specifically, we generated such a self-consistent-field potential using a 2$\times$2$\times$1 hexagonal supercell of 32 atoms containing one core-hole on the investigated element. Calculated nonresonant Cr $L_{2,3}$ XES and $2p$ XAS spectra are shown at the bottom of Figure 2, while C $K$ XES and C $1s$ XAS in Figure 3 and Ge $M_{1}$, $M_{2,3}$ XES in Figure 4. 

\subsection{Calculation of Phonon Density of States}

The supercell approach was used to compute the phonon densities of states (PhDOS) within the framework of DFT and Density-Functional Perturbation Theory (DFPT) \cite{Baroni}.
Specifically, the q-ESPRESSO software package \cite{QE} was employed to calculate real space force constants on a 2$\times$2$\times$1 supercell system. Phonon frequencies were successfully obtained from the force constants by means of the QHA code 
\cite{Isaev}. Vanderbilt ultrasoft pseudo-potentials \cite{Vanderbilt} were used for all the atomic species, while exchange-correlation effects were treated with the PBE generalized gradient approximation. Integration over the BZ was performed by using the special k-points technique \cite{Monkhorst} and a Gaussian broadening of 0.002 Ry. Electronic wave functions were plane-wave expanded with a cut-off of 70 Ry and a kinetic energy up to 700 Ry.

\begin{figure}
\includegraphics[width=87mm]{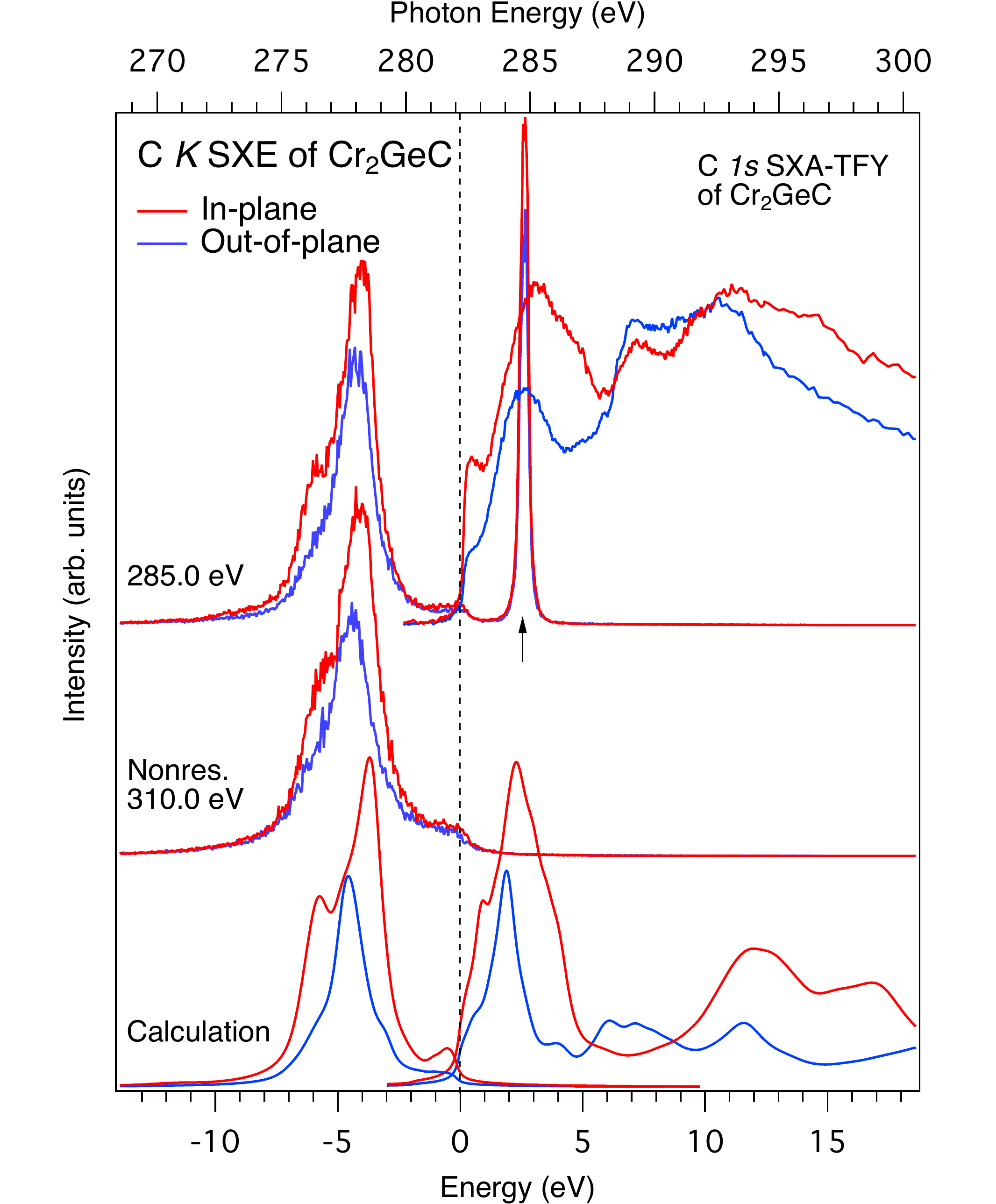} 
\vspace{0.2cm} 
\caption[] {(Color online) Top-right: C $1s$ XAS-TFY spectra of Cr$_2$GeC. Left top-to-bottom: resonant spectra excited at 285.0 (indicated by the upward arrow) and nonresonant C $K$ XES spectra at 310.0 eV. All spectra are plotted on a photon energy scale (top) and a relative energy scale (bottom) with respect to the top of the valence band. }
\label{C 1s}
\end{figure}

\section{Results}
\subsection{Cr $2p_{3/2,1/2}$ x-ray absorption and $L_{2,3}$ emission}

Figure 2 (top-right) shows XAS-TFY spectra measured at the Cr $2p$ absorption edge and normalized at 600 eV. The intensity is related to the unoccupied Cr states of Cr$_{2}$GeC that are projected via the Cr $2p$ $\rightarrow$ $3d4s$ dipole selection rules. The XAS spectra are shown as measured with the polarization vector both in the $ab$ basal-plane and along the $c$-axis of Cr$_{2}$GeC. The main XAS peak structures correspond to the $2p_{3/2}$ and $2p_{1/2}$ core levels. The observed sub-peak splittings are most salient features for the in-plane orbitals, while the peaks across the basal plane have lower intensity. The structures agree with the DFT calculations shown at the bottom-right part of Fig. 2. The branching ratios observed experimentally are different from the statistical 2:1 ratio since they are affected by both exchange as well as mixed terms between the $2p$ core states and the valence states \cite{Laskowski}. At the E$_{F}$, the intensity of the unoccupied states is 35 \% higher in the $ab$ basal laminate plane than along the $c$-axis. 

The central part of Fig. 2 shows Cr $L_{2,3}$ XES and RIXS spectra. These spectra are related to the occupied Cr $3d$ and $4s$ valence bands through the Cr $3d4s$ $\rightarrow$ $2p$ dipole selection rule of Cr$_{2}$GeC. The spectra are shown both in the $ab$ laminate plane and along the $c$-axis and were excited at two different energies at the $2p_{3/2}$ absorption maxima (indicated by vertical arrows in the XAS spectra), and also nonresonant at 620 eV. The main $L_{3}$ peak (between -2 and -5 eV below E$_{F}$) is due to the Cr $3d$ - C $2p$ orbital overlap. As observed, the in-plane spectra have higher intensities than the out-of-plane spectra. The intensity at the E$_F$,  is 40 \% larger in the $ab$ basal laminate plane compared to the intensity along the $c$-axis. The calculated spectra (Fig. 2, bottom) confirms the greater spectral intensity at the $E_{F}$ for the in-plane states (30 \%) than for the out-of-plane states, as shown in the inset. 

\subsection{C $1s$ x-ray absorption and C $K$ emission}
Figure 3 (top right) shows measured C $1s$ XAS spectra with the polarization vector both in the $ab$ basal laminate plane and along the $c-axis$, normalized at 315 eV. The intensity is related to the unoccupied C states of Cr$_{2}$GeC that are projected via the C $1s$ $\rightarrow$ $2p$ dipole selection rules. The C atoms in Cr$_{2}$GeC are located in octahedral cavities between the Cr$_{2}$C layers. Due to the apparent symmetric environment, a nearly isotropic character of the C-atoms is expected \cite{Bugnet}. However, the difference between the XAS spectra measured across and in the $ab$ basal laminate plane is relatively large. At the $E_{F}$, the data exhibit 67 \% more empty hole-like states in the \textit{ab} basal laminate plane than along the \textit{c}-axis. This difference is consistent with our calculated spectra (74 \%) at the bottom-right in Fig. 4. Differences between in- and out-of-plane also occur at higher energies above $E_{F}$. The shoulder at 0.5 eV and the peak at 6 eV that occur in-plane are more pronounced in the experiment than in the calculation. This may be due to carbon contamination on the sample or on the beamline mirrors.

The C $K$ RIXS and nonresonant XES spectra of Cr$_{2}$GeC were excited at 285.0 and 310.0 eV photon energies, respectively. Calculated nonresonant spectra are shown at the bottom of Figure 3. For the in-plane excitation, a main peak and a shoulder with $\sigma$ orbital character are clearly observed at -4.0 eV and -6.0 eV below $E_{F}$. This intensity corresponds to the C $2p_{xy}$ $\rightarrow$ $1s$ dipole transitions of Cr$_{2}$GeC. 
The spectra probed along the \textit{c}-axis instead exhibit a single peak at -4.3 eV below $E_{F}$ and are due to $2p_{z}$ states with $\pi$ orbital character. The C $2p$ peak structures hybridize mainly with the upper part of the Cr $3d$ bonding orbitals, but also with the Ge $4p$, $4s$ orbitals and the shallow $3d$ core levels. Contrary to the occupied Cr $3d$ states, the intensity at the $E_{F}$ of the occupied C $2p$ states is similar in the \textit{ab} basal laminate plane and along the \textit{c}-axis. 

\begin{figure}
\includegraphics[width=87mm]{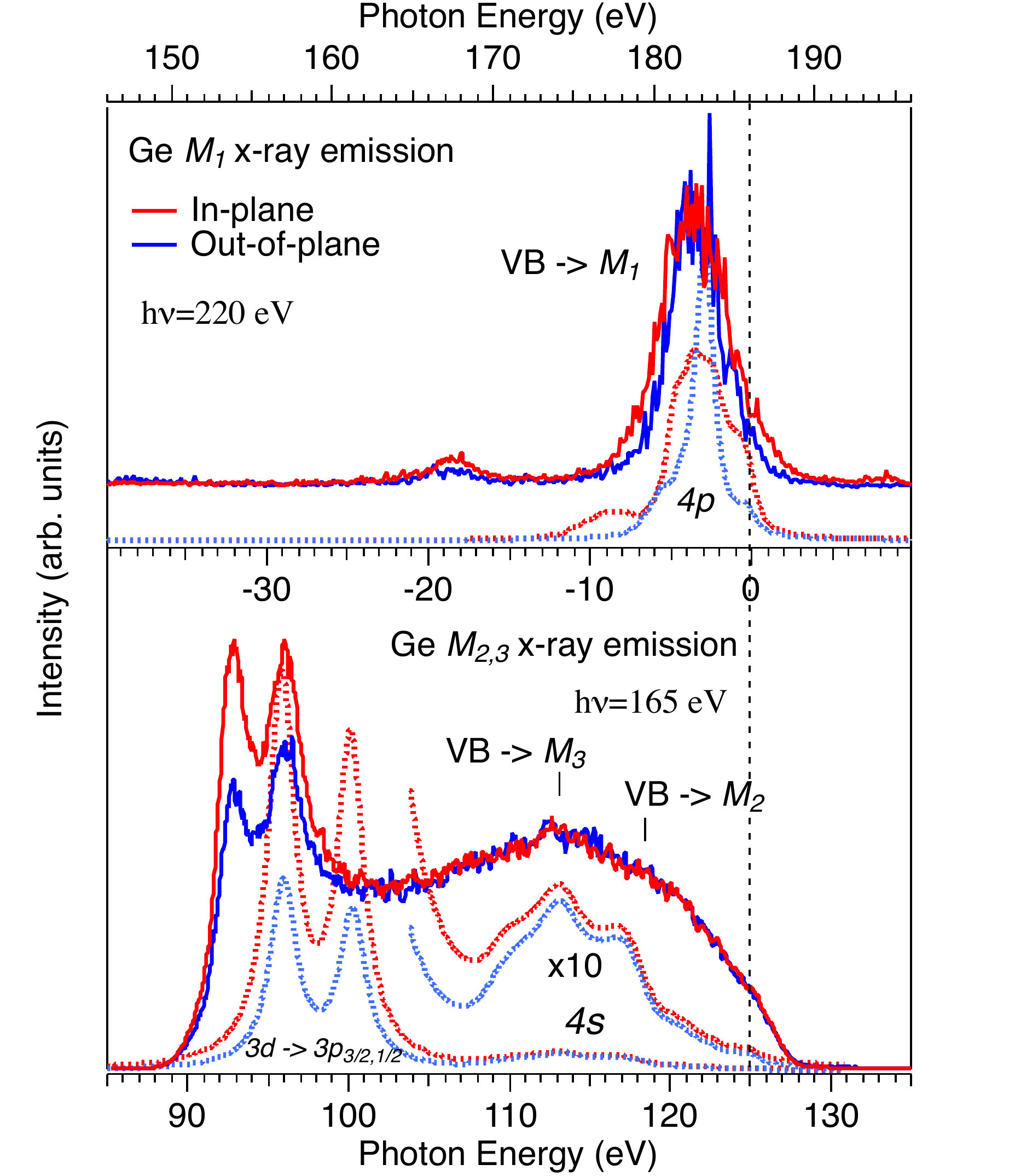} 
\vspace{0.2cm} 
\caption[] {(Color online) Ge $M_{1}$ XES spectra of Cr$_{2}$GeC 
excited at 220 eV. Bottom panel: Ge $M_{2,3}$ XES 
spectra excited at 165 eV. A common energy scale with respect to the top of the 
valence band edge (vertical dotted line) is indicated between the top and bottom 
panels. The dashed curves below the experimental data are calculated spectra for comparison.}
\label{Ge M}
\end{figure}

\subsection{Ge $M_{1}$ and $M_{2,3}$ x-ray emission}
Figure 4 (top panel) shows experimental Ge $M_{1}$ XES spectra ($4p \rightarrow 3s$ transitions) of Cr$_{2}$GeC measured nonresonantly at 220 eV photon energy probing the Ge $4p$ states across and in the laminate $ab$-plane. The spectra are plotted on a photon energy scale and relative to the $E_{F}$. Calculated Ge $M_{1}$ spectra are shown below the experimental spectra. Experimentally, the anisotropy of the Ge $4p$ states is found to be rather large with more states in the $ab$-plane than along the $c$-axis. 

The in-plane $4p_{xy}$ states with three $\sigma$ orbitals are spread out between 0 and -10 eV below $E_{F}$. On the contrary, the $4p_{z}$ states with a single bonding $\pi$ orbital are more localized at -3 eV below $E_{F}$. 
The experimental out-of-plane Ge $4p_{z}$ - $\pi$ spectrum is broader than the sharper calculated specrum, possibly due to electron-phonon interactions  with Ge displacements. At the $E_{F}$, the Ge $4p$ intensity is $\sim$10\% higher in the basal \textit{ab}-plane than along the \textit{c}-axis, both experimentally and theoretically.

The bottom panel in Fig. 4 shows experimental Ge $M_{2,3}$ XES spectra ($3d \rightarrow 3p$ and $4s \rightarrow 3p$ transitions) of Cr$_{2}$GeC aligned to the $E_{F}$ at the $3p_{1/2}$ core level. Here, we find that the isotropic $4s$-states of the Cr$_{2}$GeC valence band have an exceptionally high intensity in a broad range from -20 eV up to the E$_{F}$ in comparison to other Ge-containing Ti$_{3}$GeC$_{2}$ \cite{Magnuson2} and V$_{2}$GeC \cite{Magnuson1} compounds. Contrary to the experimental observations on Ge in Cr$_{2}$GeC, the calculated $4s$ valence band is more than 10 times weaker with most intensity in the range between -8 to -12 eV below E$_{F}$. Note that at the E$_{F}$, the intensity is two orders of magnitude higher in the experiment than in the calculated spectra. 

The shallow Ge $3d$ core levels at the bottom of the valence band between -30 and -35 eV also participate in the Cr-Ge bonding in Cr$_{2}$GeC. The measured $3d_{5/2,3/2}$ spin-orbit splitting of 3.6$\pm$0.1 eV is consistent with the difference between the $3p_{3/2,1/2}$ and $3d_{5/2,3/2}$ spin-orbit splittings observed in XPS (4.1-0.5 eV) and in Ti$_{3}$GeC$_{2}$ \cite{Magnuson2} and V$_{2}$GeC \cite{Magnuson1}. Theoretically, the spin-orbit splitting is larger (4.4 eV) and the states are also about 4.5 eV closer to the E$_{F}$ than in the experiment. The enhanced Ge $4s$ intensity observed experimentally may be attributed to electron-correlation, many-body effects or phonon vibrations, but more extensively to hybridization and charge-transfer effects, as in the case of Ga in GaN \cite{Magnuson7}.

Another interesting observation is that the intensity of the experimental Ge $3d$ states is higher in the basal \textit{ab}-plane than along the \textit{c}-axis. This is also the case for the calculated Ge $M_{2,3}$ spectra (dashed curves) with the statistical $3p_{3/2}$/$3p_{1/2}$ core-level ratio of 2:1. Note that the $M_{3}$/$M_{2}$ branching ratio is lower (0.8:1) for measurements with polarization along the $c$-axis than in the basal $ab$-plane (1:1). The trend in XES branching ratios in the transition-metal compounds is a signature of the degree of ionicity in the systems \cite{Kawai}, due to the additional Coster-Kronig process \cite{Cu,XMCD}. The lower $3p_{3/2}$/$3p_{1/2}$ peak ratio along the $c$-axis is thus an indication of higher ionicity (resistivity) than in the basal $ab$-plane. The valence-to-core matrix elements are thus found to play an important role to the spectral shape of the Ge $M_{2,3}$ XES spectra. The redistribution of intensity from the localized $3d$ states to the valence band may explain the large discrepancy between experiment and calculations for the DOS observed at E$_{F}$.

Figure 5 shows calculated Phonon Density of States (PhDOS) for the Cr, Ge and C atoms in Cr$_{2}$GeC in the $ab$-basal plane (x,y) and along the $c$-axis ($z$). Generally, the frequency in phonon spectra is proportional to the atomic mass \cite{Laeffer,Spanier}. All elements have a distinct vibration frequency along the $ab$-plane vs $c$-axis direction. For Ge, the vibration frequency along the $c$-axis is slightly larger (5.5-8 THz) than that of the basal plane (3-5 THz). The Ge atoms move preferentially within the $ab$-basal plane, while Cr and C have preference for the $c$-axis direction. For Cr, we find a $x,y$-displacement of 0.097 \AA{} and a $z$-displacement of 0.114 \AA{} from the phonon calculations. For Ge, the $x,y$-displacement is 0.128  \AA{} and the $z$-displacement 0.081 \AA{}, while for the C-atoms, the $x,y$-displacement is 0.085 \AA{} and the $z$-displacement 0.109  \AA{}. 

Rapidly moving atoms is known to change the electronic DOS at E$_{F}$ \cite{Magnuson3}. However, the relatively small magnitude of these displacements alone cannot account for the large difference in DOS between experiment and theory for Cr$_{2}$GeC. The hybridization and redistribution of charge between the shallow $3d$ core-level and the $4s$ valence band in Ge is also a necessary prerequisite for exceptionally high DOS at E$_{F}$ compared to other Ge-containing compounds. However, for the Cr atoms, the influence of phonons to the enhanced DOS at $E_{F}$ is not obvious, as the motions along the $x$, $y$ and $z$ directions are mixed.

\begin{figure}
\includegraphics[width=95mm]{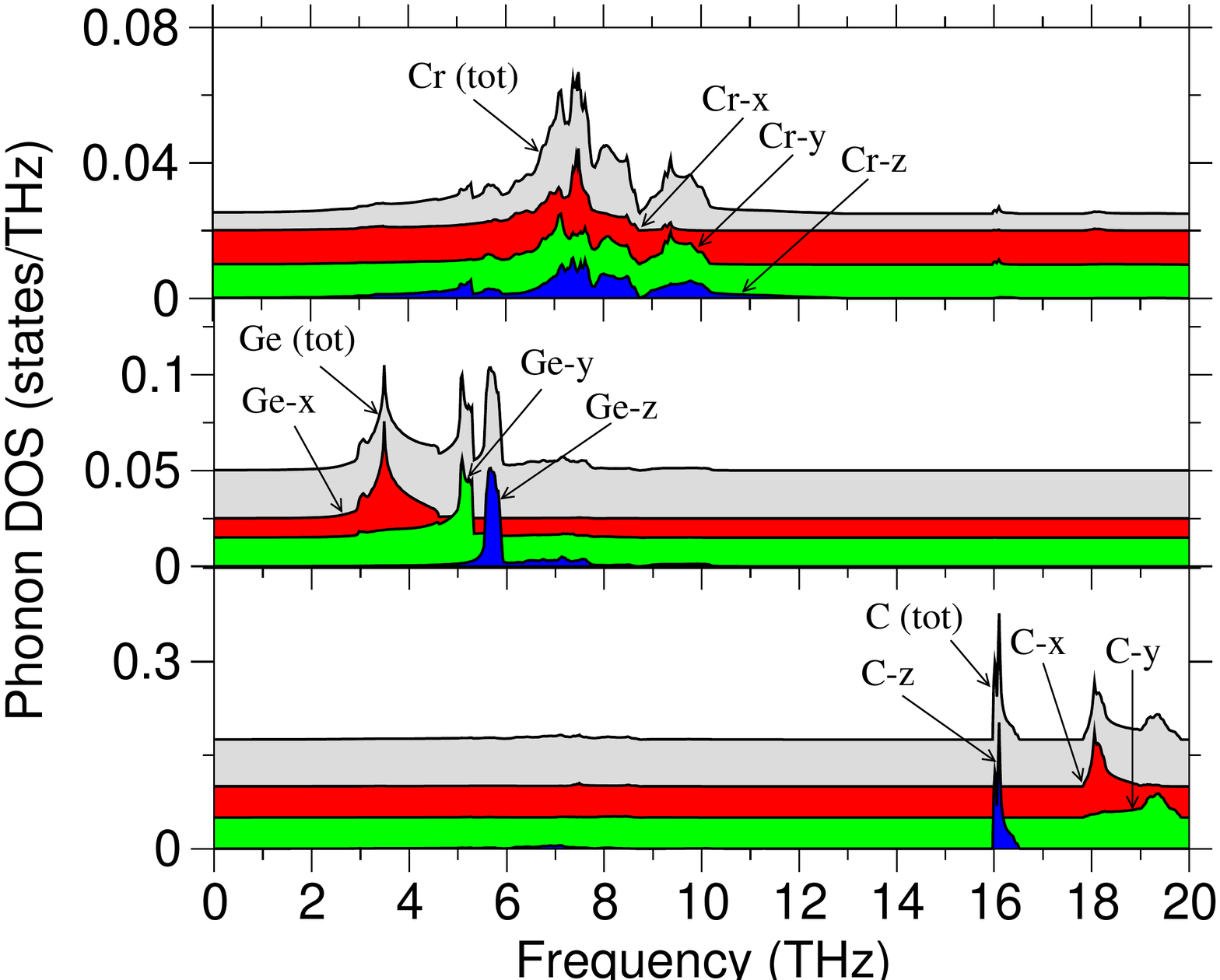} 
\vspace{0.2cm} 
\caption[] {(Color online) Calculated phonon frequency spectra of the Cr, Ge and C atoms in Cr$_{2}$GeC. The Ge atoms move preferentially within the $ab$-basal plane, while both Cr and C have more preference for the $c$-axis direction.}
\label{Phonons}
\end{figure}

\section{Discussion}
The present results indicate that the peculiar transport properties in Cr$_{2}$GeC are more complex due to the Ge-element in comparison to other $M_{n+1}AX_{n}$-phase systems where they are governed by the choice of $3d$ transition element. The observation of the dominating contribution of the $4s$ states in the Ge $M_{2,3}$ XES in Fig. 4 explains the large difference in the DOS at E$_{F}$ between experiment (22 states/eV/cell) \cite{Drulis} and DFT calculations (7.7 states/eV/cell) \cite{Maurizio}. 

In the DFT calculations, the Ge $3d$ states are overestimated and not enough intensity is redistributed to the $4s$ valence band states. The same type of deficiency in DFT calculations is known also for Ge in Ti$_{3}$GeC$_{2}$ \cite{Magnuson2} and V$_{2}$GeC \cite{Magnuson1} as well as for Ga in GaN \cite{Magnuson7}. 
We clearly observe that the calculated $4s$ states at the bottom of Fig. 4 are too low in intensity, in particular at the E$_{F}$. As this problem is a shortcoming in $3d$/$4s$ intensity ratio, we anticipate that this phenomenon is attributed to $3d$-$4s$ hybridization effects in combination with mixing of other states that change the relative population of the $3d$ and $4s$ states. 
On the other side, obtaining the DOS at $E_{F}$ by measuring the specific heat \cite{Drulis} may be sensitive to Cr$_{5}$Ge$_{3}$C$_{x}$ and Cr$_{3}$C$_{2}$ impurities. It should also be emphasized that the samples used in ref. \cite{Drulis} contained substantial amounts of Cr$_{2}$O$_{3}$ \cite{Amini} and that the fitting of the specific heat was not satisfactory above 180 K \cite{Drulis}. This may lead to an overestimated value of the actual DOS at E$_{F}$.
As our Cr$_{2}$GeC sample is phase pure with no Cr$_{5}$Ge$_{3}$C$_{x}$ impurities \cite{Eklund2}, we only consider the very small fraction of Cr$_{3}$C$_{2}$ phase that could contribute to an increased DOS at E$_{F}$ \cite{Cr3C2}.
However, if there is an impurity phase that causes the high DOS at E$_{F}$, this impurity should contain Ge and that excludes Cr$_{3}$C$_{2}$.

The Cr and Ge atoms are also known to have a correlated motion where the Cr-Ge bond length is kept constant \cite{Nina}. The rattling Ge atoms thus drive the Cr atoms in the structure. The Cr $3d$ DOS have an intensity at E$_{F}$ that is sensitive to the particular in-plane and out-of-plane phonon modes of the Cr atoms. However, the change in DOS at the E$_{F}$ is not only due to the movement of the Cr atoms, but also to the moving Ge and C atoms where the Ge atoms move preferentially within the $ab$-basal plane, while both Cr and C have more preference for the $c$-axis direction. 

The lowest-frequency vibrations are associated with the Ge atoms in Cr$_{2}$GeC, similarly to the case of Al in Cr$_{2}$AlC \cite{Bugnet} and Ti$_{3}$AC$_{2}$ (A =Si, Al, and Ge), where the lowest-energy vibrations are from the A element \cite{Togo, Magnuson2,Magnuson3}. This finding is consistent with the fact that the Cr-C bonding type is the stiffest. This is also in agreement with what was recently found in Cr$_{2}$AlC \cite{Bugnet}. However, the Ge and Cr vibrations in Cr$_{2}$GeC appear to be more decoupled than in the case of Al and Cr in Cr$_{2}$AlC \cite{Bugnet}. This type of tangled phonon scenario in MAX-phases will likely influence the core-level spectra at finite temperature values as well.  

The electronic structure anisotropy of Cr$_{2}$AlC has been studied previously by electron energy loss spectroscopy (EELS) at the C-$K$ edge and by polarized XAS at the Al-K edge  \cite{Bugnet}. A marked anisotropy at the Al-site was observed, in addition to a much smaller anisotropic character of the C-site, essentially attributed to the octahedral environment of the C-sites in the M$_{n+1}$AX$_{n}$-phase structure, which is much more isotropic than the trigonal prismatic site occupied by Al-atoms. The relatively small anisotropy of the C-$K$ edge was exacerbated by the average of crystallographic orientations probed in EELS, as opposed to polarized x-ray beams. Here, for Cr$_{2}$GeC, the anisotropy of the C-$K$ edge is present in the XAS and XES data presented in Fig. 3, although it may also be somewhat influenced by the presence of amorphous C, resulting from contamination of the sample. The anisotropy is consistent with the case of other M$_{n+1}$AX$_{n}$ - phase systems such as V$_{2}$GeC \cite{Magnuson1} and Ti$_{3}$SiC$_{2}$ \cite{Magnuson3}.
Although the local symmetry at each atomic site is important, the distribution of the electronic states also obeys the whole $k$-space topology of the M$_{n+1}$AX$_{n}$-phase unit cell, that is not highly symmetric. This is clearly observed at, for instance, the $c$-axial elongation of the unit cell.

Our results clearly show that the anisotropy in the electronic structure and chemical bonding of Cr$_{2}$GeC is substantial for the Cr $3d$, C $2p$ and Ge $3d$ states. The generally greater amount of both occupied and unoccupied Cr $3d_{xy}$, $3d_{x^2-y^2}$ and C $2p_{xy}$ electronic states in the basal $ab$-plane than Cr $3d_{z^2}$ and C $2p_{z}$ states along the $c$-axis should contribute to higher electrical conductivity in the basal $ab$-plane for a large range of temperatures \cite{Eklund2}. However, since the C $2p$ states are in minority they should not affect the conductivity as much as the Cr $3d$ states. 

The calculations indicate that the Ge $4p_{xy}$-$\sigma$ states of the basal $ab$-plane are spread out in the region 0 to -10 eV, while the more localized single out-of-plane $4p_{z}$-$\pi$ states along the $c$-axis have a peak maximum around -3 eV below $E_{F}$. Contrary to the case of the phonon rattling of the Si $3d$ states in Ti$_{3}$SiC$_{2}$ \cite{Magnuson3}, the phonon modes of the isotropic Ge $4s$ states in Cr$_{2}$GeC should not affect the anisotropy of the transport properties as much as the $3d$ states of the transition element. 
The differences in intensity of the Ge $4s$ band explain the large difference in the DOS at $E_{F}$ between experiment and theory.

The strong covalent Cr $3d$ - C $2p$ bonding in Cr$_{2}$GeC occurs in the hybridization region between -2 to -5 eV below $E_{F}$ and the shallow Ge $3d$ core level states are located at -30 eV below the Fermi level. Intercalation of Ge monolayers into CrC with relatively weak Cr $3d$ - Ge $4p$ bonds thus implies that the Cr-C bonds are strengthened. This also affects the Cr-Cr bonds that are strengthened, which enhances the metallic character and the density of states at the Fermi level of Cr$_{2}$GeC. Contrary to the elastic and stiffness properties, the transport properties depend more clearly on the states at the E$_{F}$. 

The fact that the number of electronic states at E$_{F}$ are larger experimentally than those computed with DFT, is due to the additional electronic states provided to the E$_{F}$ by the Ge $4s$ states, which exhibit high intensity in the Ge $4s$ XES. The Ge $4s$ and Ge $3p$ states hybridize more with the Cr $3d$ states than DFT predicts, transferring more electron density towards the Fermi level. This effect is clearly observed in the $3p$/$4s$ intensity branching ratio in Fig. 4. Experimentally, the intensity ratio between the $3p\rightarrow3d$ and the $4s\rightarrow3d$ peaks is much lower than the one computed with DFT, as a consequence of the charge transfer from the shallow Ge $3p$ core levels to the Ge $4s$ states. This hybridization effect clearly manifests on the Ge $M_{2,3}$ XES (i.e., the partial DOS) and probably also has an important effect on the total DOS at E$_{F}$. However, we cannot exclude that impurities can also contribute to the very large experimentally observed DOS at E$_{F}$ \cite{Drulis}. Furthermore, not only the crystal structure and the constituent elements, but also the intrinsic rattling motion of the Ge atoms, can increase the density of states at $E_F$ and hence influence the materials transport properties. 
While the observed intensity increase of the Ge $4s$ states is isotropic at the $E_F$, the difference in the phonon frequencies in the basal $ab$-plane and along the $c$-axis should affect the conductivity in Cr$_{2}$GeC.

\vspace{12pt}
\section{Conclusions}

Polarization-dependent experimental and theoretical results provide understanding of what electronic states affect the anisotropy of the electronic structure, chemical bonding, and hybridization regions in Cr$_{2}$GeC. The combination of hybridization, charge-transfer redistribution, Cr $3d$ electron correlation effects and electron-phonon coupling give rise to the observed transport properties. The occupied electronic DOS in the basal $ab$-plane is larger than those along the $c$-axis for the Cr $3d$ and Ge $4p$ states. The Ge $4s$ and shallow Ge $3p$ core-level states exhibit stronger hybridization with the Cr $3d$ states than DFT predicts in the one-electron approximation, transferring extra electron density towards the E$_{F}$. Thus, as shown in this work, the large intensity difference between the experimentally observed Ge $4s$ valence band states and the $3d$ states essentially explains the large difference in the experimental and calculated density of states at the Fermi level.

\section{Acknowledgements}

We thank the staff at MAX IV Laboratory for experimental support as well as M. Jaouen and V. Mauchamp for discussions. 
This work was financially supported by the Swedish Research Council Linnaeus Grant LiLi-NFM and the Swedish
Foundation for Strategic Research (SSF) through the Synergy Grant FUNCASE and
Future Research Leaders 5 program. P. E. also acknowledges the University of Poitiers for a Visiting Professor position.
M. Mattesini acknowledges financial support by the Spanish Ministry of Economy and Competitiveness (CGL 2013-41860-P) and by the BBVA Foundation.


\begin{thebibliography}{100}


\bibitem{Wang}  J. Wang and Y. Zhou; Annu. Rev. Mater. Res. \textbf{39}, 415-443 (2009).

\bibitem{Eklund1} P. Eklund, M. Beckers, U. Jansson, H. H\"{o}berg, L. Hultman; Thin Solid Films 
\textbf{518}, 1851-1878 (2010).

\bibitem{Barsoum1} M. W. Barsoum; Prog. Solid State Chem. \textbf{28}, 201-281 (2000).

\bibitem{Jeitschko} W. Jeitschko \textit{et al.}; Monatshefte Chem. \textbf{94}, 844 (1963).

\bibitem{Bouhemadou} A. Bouhemadou; Appl. Phys. A, \textbf{96}, 959 (2009).

\bibitem{Scabarozi} T. H. Scabarozi, S. Amini, O. Leaffer, A. Ganguly, S. Gupta, W. Tambussi, 
S. Clipper, J. E. Spanier, M. W. Barsoum, J. D. Hettinger and S. E. Lofland; 
J. Appl. Phys. \textbf{105}, 013543 (2009).

\bibitem{Cabioch} T. Cabioch, P. Eklund, V. Mauchamp, M. Jaouen, M. W. Barsoum; 
J. Eur. Cer. Soc. \textbf{33}, 897 (2013).

\bibitem{Nina}  N. J. Lane, S. Vogel, and M. W. Barsoum; J. Am. Ceramic Soc. \textbf{94}, 3473 (2011).

\bibitem{Zhou} W. Zhou, L. Liu and P. Wu; J. Appl. Phys. \textbf{106}, 033501 (2009).

\bibitem{Ramzan1} M. Ramzan, S. Leb\'egue and R. Ahuja; Phys. Status Solidi RRL \textbf{5}, 122 (2011).

\bibitem{Ramzan2} M. Ramzan, S. Lebegue and R. Ahuja; Solid State Commun. \textbf{152} 1147 (2012).

\bibitem{Maurizio} M. Mattesini and M. Magnuson; J. Phys.: Condens. Matter \textbf{25}, 035601 (2013).

\bibitem{Dahlqvist} M. Dahlqvist, B. Alling and J. Rosen; J. Appl. Phys. \textbf{113}, 216103 (2013).

\bibitem{Jaouen} M. Jaouen, M. Bugnet, N. Jaouen, P. Ohresser, V. Mauchamp,
T. Cabioc’h and A. Rogalev; J. Phys. Cond. Mat. \textbf{26}, 176002 (2014).

\bibitem{Barsoum2} M. Barsoum, T. H. Scabarozi, S. Amini, J. D. Hettinger and S. E. Lofland; J. Am. Ceram. Soc. \textbf{94}, 4123 (2011).

\bibitem{Drulis} M. K. Drulis, H. Drulis, A. E. Hackemer, O. Leaffer, J. Spanier, S. Amini, M. W. Barsoum, T. Guilbert and T. El-Raghy; J. Appl. Phys. \textbf{104}, 023526 (2008).

\bibitem{Magnuson4} M. Magnuson, M. Mattesini, S. Li, C. H\"{o}glund, M. Beckers, L. Hultman and O. Eriksson;
Phys. Rev. B., \textbf{76}, 195127 (2007). 

\bibitem{Magnuson5} M. Magnuson, O. Wilhelmsson, J. -P. Palmquist, U. Jansson, M. Mattesini, S. Li, R. Ahuja and O. Eriksson;
Phys. Rev. B. \textbf{74}, 195108 (2006). 

\bibitem{Magnuson6} M. Magnuson, M. Mattesini, O. Wilhelmsson, J. Emmerlich, J. -P. Palmquist, S. Li, R. Ahuja, L. Hultman, O. Eriksson and U. Jansson;
Phys. Rev. B. \textbf{74} , 205102 (2006).

\bibitem{Eklund2} P. Eklund, M. Bugnet, V. Mauchamp, S. Dubois, C. Tromas, J. Jensen, L. Piraux, L. Gence, M. Jaouen and T. Cabioh; Phys Rev. B \textbf{84}, 075424 (2011).

\bibitem{MAX-lab} M. Magnuson, M. Mattesini, C. H\"{o}glund, I. A. Abrikosov, J. Birch and L. Hultman,
Phys. Rev. B \textbf{78}, 235102 (2008). 

\bibitem{Nanocomp}  M. Magnuson, E. Lewin, U. Hultman and U. Jansson; Phys. Rev. B \textbf{80}, 
235108 (2009).

\bibitem{Klementev} K. V. Klementev; J. Phys. D: Appl. Phys. 34, 209-17 (2001);
XANES dactyloscope for Windows, K. V. Klementiev, freeware: www.desy.de/~klmn/xanda.html
www.cells.es/old/Beamlines/CLAESS/software/

\bibitem{Cu} M. Magnuson, N. Wassdahl, and J. Nordgren, Phys. Rev. B \textbf{56}, 12238 (1997).

\bibitem{Kohn}  W. Kohn and L. J. Sham; Phys. Rev. \textbf{140}, A1133-A1138 (1965).

\bibitem{vonBarth} U. von Barth and G. Grossmann; Phys. Rev. B \textbf{25}, 5150-5179 (1982).

\bibitem{Blaha} Blaha P, Schwarz K, Madsen G K H, Kvasnicka D and Luitz J, (2001) WIEN2K,
\textit{An Augmented Plane Wave + Local Orbitals Program for Calculating Crystal}
\textit{Properties} (Karlheinz Schwarz, Techn. Universität Wien, Austria). ISBN 
3-9501031-1-2.

\bibitem{Perdew} J. P. Perdew, S. Burke, and M. Ernzerhof; Phys. Rev. Lett., \textbf{77} 3865 (1996).

\bibitem{LaCoO3} M. Magnuson, S. M. Butorin, C. S\aa{}the, J. Nordgren and P. Ravindran;
Europhys. Lett.  \textbf{68}, 289 (2004). 

\bibitem{Baroni} S. Baroni, S. de Gironcoli, A. Dal Corso and P. Giannozzi; Rev. Mod. Phys. \textbf{73}, 515 (2001).

\bibitem{QE} Quantum-ESPRESSO, see http://www.quantum-espresso.org and http://www.pwscf.org.

\bibitem{Isaev} E. Isaev, QHA project, http://qe-forge.org/qha. n.d.

\bibitem{Vanderbilt} D. Vanderbilt; Phys. Rev. B \textbf{41} 7892 (1990).

\bibitem{Monkhorst} H. J. Monkhorst and J. D. Pack; Phys. Rev. B \textbf{13}, 5188 (1976).

\bibitem{Laskowski} R. Laskowski and P. Blaha; Phys. Rev. B \textbf{82}, 205104-205110 (2010).

\bibitem{Bugnet} M. Bugnet, M. Jaouen, V. Mauchamp, T. Cabioc’h and G. Hug; Phys. Rev. B \textbf{90}, 195116 (2014). 

\bibitem{Magnuson2} M. Magnuson, J.-P. Palmquist, M. Mattesini, S. Li, R. Ahuja, O. Eriksson, J. Emmerlich, O. Wilhelmsson, P. Eklund, H. H\"{o}gberg, L. Hultman, and U. Jansson; Phys. Rev. B \textbf{72}, 245101-245110 (2005).

\bibitem{Magnuson1} M. Magnuson, O. Wilhelmsson, M. Mattesini, S. Li, R. Ahuja, O. Eriksson, H. H\"{o}gberg, L. Hultman, and U. Jansson; Phys. Rev. B \textbf{78}, 035117 (2008).

\bibitem{Magnuson7} M. Magnuson, M. Mattesini, C. H\"{o}glund, J. Birch, and L. Hultman;
Phys. Rev. B \textbf{81}, 085125 (2010).

\bibitem{Kawai} J. Kawai, K. Nakajima and Y. Gohshi, Spectrochem. Acta, Part B \textbf{48}, 1281 (1993).

\bibitem{XMCD} M. Magnuson, L.-C. Duda, S. M. Butorin, P. Kuiper, and J. Nordgren, Phys. Rev. B \textbf{74}, 172409 (2006).

\bibitem{Laeffer} O. D. Leaffer, S. Gupta, M. W. Barsoum, and J. E. Spaniera; 
J. Mater. Res. \textbf{22}, 2651 (2007).

\bibitem{Spanier} J. E. Spanier, S. Gupta, M. Amer and M. W. Barsoum; 
Phys. Rev. B \textbf{71}, 012103 (2005).

\bibitem{Magnuson3}  M. Magnuson, M. Mattesini, N. Van Nong, P. Eklund and  L. Hultman; Phys. Rev. B \textbf{85}, 195134 (2012).

\bibitem{Amini} S. Amini, A. Zhou, S. Gupta, A. DeVillier, P. Finkel, and M. W. Barsoum; J. Mater. Res. \textbf{23}, 2157 (2008).

\bibitem{Cr3C2} Y. F. Li, Y. M. Gao, B. Xiao, T. Min, Y. Yang, S. Ma, D. W. Yi, J. Allys and Comp. \textbf{509}, 5242 (2011).

\bibitem{Togo} A. Togo, L. Chaput, I. Tanaka, and G. Hug, Phys. Rev. B \textbf{81}, 174301 (2010).


\end{thebibliography}
\end{document}